\documentstyle[prd,aps,epsfig]{revtex}

\newcommand{\be}{\begin{eqnarray}}
\newcommand{\ee}{\end{eqnarray}}
\newcommand{\nn}{\nonumber}
\renewcommand{\d}{{\mathrm d}}
\renewcommand{\i}{{\mathrm i}}
\newcommand{\e}{{\mathrm e}}
\newcommand{\feynmandag}{\slash\hspace{-0.55em}}
\newcommand{\alphas}{\alpha_{{\mathrm s}}}
\newcommand{\xBj}{x_{{\mathrm Bj}}}
\newcommand{\mpp}{m_{\pi\pi}}

\begin{document}
\draft
\title{Angular distributions in hard exclusive production of pion pairs}
\author{B.~Lehmann-Dronke, A.~Sch\"afer}
\address{Institut f\"ur Theoretische Physik, Universit\"at Regensburg,
D-93040 Regensburg, Germany}
\author{M.~V.~Polyakov$^{a,b}$, K.~Goeke$^b$}
\address{$^a$ Petersburg Nuclear Physics Institute, Gatchina, 188350 Russia}
\address{$^b$ Institut f\"ur Theoretische Physik II, Ruhr-Universit\"at
Bochum, D-44780 Bochum, Germany}
%\date{\today}
\maketitle
\begin{abstract}
Using the leading order amplitudes of hard exclusive electroproduction
of pion pairs we have analyzed the angular distribution of the two
produced particles. At leading twist a pion pair can be produced only
in an isovector or an isoscalar state. We show that certain components
of the angular distribution only get contributions from the
interference of the $I=1$ and the (much smaller) $I=0$
amplitude. Therefore our predictions prove to be a good probe of
isospin zero pion pair production. We predict effects of a measurable size
that could be observed at experiments like HERMES.
We also discuss how hard exclusive pion pair production
can provide us with new information on the effective chiral Lagrangian.
\end{abstract}

\section{Introduction}

The factorization theorem \cite{CFS} states that
the amplitude of the exclusive production process
\be
\gamma^\ast_L + T \to F + T'
\label{processgeneral}
\ee
at large invariant collision energy $\sqrt{s}\to\infty$, large
virtuality of the (longitudinally polarized) photon $Q^2\to\infty$,
fixed $Q^2/s$ (Bjorken limit), and with
$-t,m_T^2,m_{T'}^2,m_F^2\ll Q^2$ can be written in the form
\be
\sum _{i,j} \int\d z\d x_1 f_{i/T}^{T'}(x_1,x_1-\xBj,t)\,
H_{ij}\Big(\frac{x_1}{\xBj},Q^{2},z\Big)\, \Phi^F_j(z)
+\mbox{power-suppressed corrections}\,,\label{factorization}
\ee
where $H_{ij}$ is a hard part computable in pQCD as a series in
$\alphas$, $\Phi^F_j$ is the distribution amplitude of the
hadronic state $F$, and $f_{i/T}^{T'}$ is a $T\to T'$ skewed
parton distribution \cite{CFS,Dmuller,Ji,Rad} (for a review see
\cite{JiReview}). Skewed parton distributions (SPD's) are related
to matrix elements of bilocal operators between states of
different momentum. They are generalizations of the usual parton
distributions which parameterize the diagonal matrix elements of
the corresponding operators. Therefore hard exclusive
electroproduction opens a new way to study the partonic structure
of hadrons. Naturally, for the experimentally accessible range of
$Q^2$ the size of the $1/Q$ suppressed power corrections can be
rather large, the quantitative estimates of higher twist
corrections are therefore in the focus of recent investigation
\cite{HT}. Still, the formalism has been used to investigate
exclusive production of single light mesons
\cite{CFS,Brodsky:1994kf,Radrho,MPW,Vander}. For
the starting point of the present work it is important to note
that the factorization theorem is not limited to the case where
the produced hadronic state $F$ in reaction (\ref{processgeneral})
is a one particle state provided that its invariant mass is small
compared to $Q$ \cite{Freund:2000xg}.

In the present paper we present a detailed analysis of hard exclusive
electroproduction of two pions to leading order in the strong coupling
constant $\alphas$. In that case $\Phi^F_j$ in the expression
(\ref{factorization}) is a two-pion distribution amplitude
($2\pi$DA). At leading twist the pion pair can be produced in a state
with isospin one or zero. For $I=1$ the process is dominated by the
$\rho$ peak. (So in a certain sense our approach can be considered as
an alternative description of $\rho$ production, which has been studied
earlier using a distribution amplitude for the $\rho$ meson
\cite{CFS,Brodsky:1994kf,Radrho,MPW,Vander}). The description of pion pair
production at small $\xBj$ in terms of $2\pi$DA's was considered in
Ref.~\cite{CP}, where it was demonstrated that such a description
allows us to determine the details of the partonic structure
of the pion and the $\rho$ meson from data on the di-pion mass
distributions in hard diffraction alone.

Here we are mainly interested in the production of $I=0$ pion
pairs. It seems to be difficult to investigate isoscalar pair
production by the measurement of total cross sections because identifying
$\pi^0\pi^0$ production  events would require detecting 4 correlated
photons whereas $\pi^+\pi^-$ production is dominated by the isovector
contribution due to the $\rho$ resonance (see our analysis in
\cite{maxben}). However, the angular distribution of the
produced pion pairs contains components that depend only on the
interference of the $I=0$ and the $I=1$ channel
\cite{maxben,DGP}. Using specific
models for the involved SPD's and $2\pi$DA's we predict effects of a
measurable size and suggest the most favorable kinematic range for
such measurements.

The interesting feature of hard exclusive production of pion pairs
in an isoscalar state is that the pions are produced
not only by a collinear $q\bar q$ pair but also
(and in the same order in $1/Q$ and $\alphas$) by two
collinear gluons. This observation opens a possibility to
access the gluon content of the isoscalar $\pi\pi$ states.
In the analysis of the present work we shall assume that
the elusive isoscalar $\pi\pi$ resonance $f_0(400-1200)$ has no
anomalously large gluon contents as it was suggested e.g.\ in
\cite{MO,N}.
The use of hard exclusive pion pair production
for ``gluonometry'' of the low-lying isoscalar resonances will be
considered elsewhere.

\section{Amplitudes and intensity densities}\label{section2}

In this section we calculate the amplitude of hard two-pion
electroproduction to leading order in $1/Q^2$ and $\alphas$
in terms of SPD's and $2\pi$DA's and show how they are connected to
the so called intensity densities which we define as
\be
\langle P_l(\cos\theta)\rangle^{\pi\pi}:=
\frac{\int_{-1}^1\d\cos\theta\,P_l(\cos\theta)\,
\frac{\d\sigma^{\pi\pi}}{\d\cos\theta}}
{\int_{-1}^1\d\cos\theta\,\frac{\d\sigma^{\pi\pi}}{\d\cos\theta}}\,.
\label{IDgeneral}
\ee
\begin{figure}
\centering
\epsfig{file=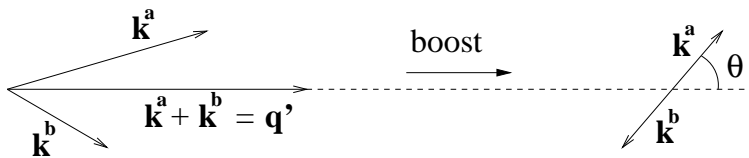, width=10cm}
\caption{Definition of the scattering angle $\theta$.}
\label{kinematics}
\end{figure}
Here $\theta$ is defined as the scattering angle of the $\pi^+$ or
one $\pi^0$ (for $\pi^+\pi^-$ or $\pi^0\pi^0$ production
respectively) in the center of mass system of the two pions with
respect to their total momentum e.g.\ in the laboratory system
(see Fig.~\ref{kinematics})\footnote{In
  principle one can analogously define a more general intensity density
  $\langle Y_{lm}(\theta,\phi)\rangle^{\pi\pi}$. However, our analysis
  shows that at leading twist the production rate is independent of
  the azimuthal angle $\phi$. Therefore non-vanishing results are
  obtained only for the intensity densities $\langle
  Y_{l0}(\theta,\phi)\rangle^{\pi\pi}$, which are up to a constant
  normalization factor equal to $\langle
  P_l(\cos\theta)\rangle^{\pi\pi}$
  defined in (\ref{IDgeneral}).}.

We consider the process
\be
\gamma_L^*(q) + B_1(p)\to \pi\pi(q')+B_2(p+\Delta)\,,
\label{subprocess}
\ee
where a linear polarized virtual photon with momentum $q$ and
a nucleon $B_1$ with momentum $p$ produce a final state with a baryon
$B_2$ (a nucleon or a heavier state) with momentum
$p'=p+\Delta$ and a pion pair with total momentum $q'$.
Note that the amplitude for the corresponding production process
with a transversally polarized virtual photon is $1/Q$ power
suppressed (with $Q^2=-q^2$). Therefore the (sub-) process
(\ref{subprocess}) with a longitudinally polarized photon is the
only leading contribution to the electroproduction reaction
$e^-(l)+B_1(p)\to \pi\pi(q')+e^-(l-q)+B_2(p+\Delta)$.
Using two light cone vectors $n$ and $\tilde n$ normalized such that
\be
n\cdot (p + p') = 2\,,\hspace{1cm}\tilde n\cdot n = 1
\ee
the relevant momenta can be expressed as
\be
p^\mu&=&(1+\xi)\tilde n^\mu+
(1-\xi )\frac{\bar m^2}{2} n^\mu -\frac 12 \Delta_\perp^\mu
\nn\\
p^{\prime \mu}&=&(1- \xi)\tilde n^\mu+
(1+ \xi)\frac{\bar m^2}{2} n^\mu +\frac 12 \Delta_\perp^\mu
\nn\\
q^{\mu}&=&-2 \tilde\xi \tilde n^\mu+
\frac{ Q^2}{4\tilde\xi} n^\mu
\ee
with
\be
\bar m^2&=&\frac 12
\Bigl(m_{B_1}^2+m_{B_2}^2-\frac{t}{2} \Bigr)
\nn\\
\tilde \xi&=&\xi\Bigl(1-\frac{1}{Q^2}(q'^2-t-4\xi^2\bar
m^2)\Bigr)
+{\cal O}\Big(\frac{1}{Q^4}\Big)\,.
\ee
$\Delta_\perp$ is defined to be that spacelike component of the momentum
transfer $\Delta$ which is transverse to the light cone vectors $n$ and
$\tilde n$, $t$ is the momentum transfer squared $t=\Delta^2$, and
$\xi$ is the skewedness parameter describing the longitudinal
component of the momentum transfer
\be
\xi = -\frac{1}{2}(n\cdot \Delta)\,,
\ee
which in the Bjorken limit $Q^2\to\infty$ can be expressed in terms of
the Bjorken variable $\xBj=\frac{Q^2}{2p\cdot q}$ as
\be
\xi= \frac{\xBj}{2-\xBj}\,.
\ee
\begin{figure}
\centering
\epsfig{file=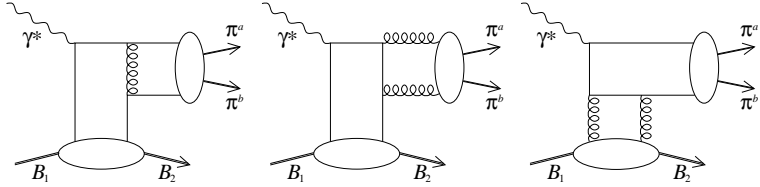, width=10cm}
\caption{Examples of the three classes of
leading order diagrams for hard exclusive pion pair
production.}
\label{feynmandiagrams}
\end{figure}

The leading order amplitude for the reaction (\ref{subprocess})
corresponds to diagrams of the type shown in
Fig.~\ref{feynmandiagrams} and has the form (see also \cite{maxben,DGP})
\be
T^{\pi^a\pi^b}&=&\langle B_2(p'), \pi^a\pi^b(q')| J^{\mathrm e.m.}\cdot
\varepsilon_L|B_1(p)\rangle
\nn\\
&=&-(e 4 \pi \alpha_s)\frac{1}{2 Q}
\int_{-1}^1 \d\tau \int_0^1 \d z
\nn\\
&&\times\Biggl\{\frac{C_F}{N_c}\sum_{f,f'}
F_{ff'}(\tau,\xi,t) \Phi_{f'f}^{ab}(z,\zeta,\mpp)
\biggl[\frac{e_{f'}}{z(\tau+\xi-\i\epsilon)}+
\frac{e_f}{(1-z)(\tau-\xi+\i\epsilon)}
\biggr]
\nn\\
&&\;\;-\frac{1}{N_c}\sum_f e_f
F_{ff}(\tau,\xi,t) \Phi_G^{ab}(z,\zeta,\mpp)
\frac{1}{z(1-z)}
\biggl[\frac{1}{\tau+\xi-\i\epsilon}-
\frac{1}{\tau-\xi+\i\epsilon}
\biggr]
\nn\\
&&\;\;+\frac{1}{2N_c}\sum_f e_f
F_G(\tau,\xi,t) \Phi_{ff}^{ab}(z,\zeta,\mpp)
\frac{1}{z(1-z)}
\biggl[\frac{1}{\tau+\xi-\i\epsilon}+
\frac{1}{\tau-\xi+\i\epsilon}
\biggr]
\Biggr\}\,,
\label{Tmatrix}
\ee
where $e_f$ is the electric
charge of a quark of flavor $f$ in units of the proton charge $e$ and
$\varepsilon_L^\mu$ is the longitudinal photon polarization vector
\be
\varepsilon_L^\mu=\frac{2\tilde\xi}{Q}\tilde n^\mu+\frac{Q}{4\tilde\xi}n^\mu
\ee
($\Rightarrow\varepsilon_L^\mu=\frac{1}{Q}(q^3,0,0,q^0)$ in a frame with
$n^\mu\propto(1,0,0,1)$ and $\tilde n^\mu\propto(1,0,0,-1)$).
The functions $F_{ff'}(\tau,\xi,t)$ and $F_G(\tau,\xi,t)$ are skewed
parton distributions for quarks or gluons respectively defined as
\be
F_{ff'}(\tau,\xi,t)&=&
\int\frac{\d\lambda}{2\pi}e^{\i\lambda \tau}
\langle B_2(p^{\prime })|T \Big[
\bar \psi_{f'}\!\Big(\!-\!\frac{\lambda n}{2}\Big){\feynmandag n}
\psi_f\!\Big(\frac{\lambda n}{2}\Big)\Big]
|B_1(p)\rangle
\label{SPDq}
\\
F_G(\tau,\xi,t)&=&
\frac{1}{\tau}\int\frac{\d\lambda}{2\pi}e^{\i\lambda \tau}n^\mu n^\nu
\langle B_2(p^{\prime })|T \Big[
F_{\mu}^{A\alpha}\!\Big(\!-\!\frac{\lambda n}{2}\Big)
F^{A}_{\alpha\nu}\!\Big(\frac{\lambda n}{2}\Big)\Big]
|B_1(p)\rangle\,.\label{SPDg}
\ee
The gluon color index $A$ is understood to be summed over.
The two-pion distribution amplitudes
\be
\Phi_{f'f}^{ab}(z,\zeta,\mpp)&:=&
\int\frac{\d\lambda}{2\pi} e^{-\i\lambda z(q'\cdot\tilde n)}
\langle\pi^a\pi^b|
T\big[\bar\psi_f(\lambda\tilde n){\feynmandag\tilde n}
\psi_{f'}(0)\big]
|0\rangle
\label{2piDAq}
\ee
and
\be
\Phi^{ab}_G (z,\zeta,\mpp)&:=&
\frac{1}{\tilde n\cdot q'}
\int\frac{\d\lambda}{2\pi} e^{-\i\lambda z (q'\cdot\tilde n)}
\tilde n^\mu\tilde n^\nu
\langle\pi^a\pi^b|
T\big[
F_{\mu}^{A\alpha}(\lambda\tilde n)
F^{A}_{\alpha\nu}(0)\big]
|0\rangle
\label{2piDAg}
\ee
describe how a pion pair is produced by two quarks or gluons
respectively. The variable $z$ is the fraction
of the longitudinal (along the vector $n$) meson pair momentum
$q'=q-\Delta$ carried by one of the two partons. The variable $\zeta$
characterizes the distribution of the longitudinal component of $q'$
between the two pions and is given in terms of the momentum $k^a$
of the pion $\pi^a$ and the total momentum of the pion pair $q'$ by
\be
\zeta=\frac{k^a\cdot\tilde n}{q'\cdot\tilde n}\,.
\ee
Let us note that $\zeta$ is related to the angle $\theta$ defined
above by
\be
\beta \cos \theta =2\zeta-1 \quad{\mathrm with}\quad
\beta:=\sqrt{1-\frac{4 m_\pi^2}{\mpp^2}}\,.
\ee
In the definitions (\ref{SPDq}) -- (\ref{2piDAg}) a gauge link of the form
${\cal P}[\exp(\i g\int_{z_2}^{z_1}A_\mu\d z^\mu)]$ with
$A_\mu=\sum_A t^A A^A_\mu$ is implied to be inserted between the two
operators at different space--time coordinates. Also we do not write out
explicitly the scale dependence of the SPD's and $2\pi$DA's.

In the following we will restrict ourselves to the case where
$B_1$ and $B_2$ both are proton states. Then the pion pair can be
either $\pi^+\pi^-$ or $\pi^0\pi^0$. The $2\pi$DA's (\ref{2piDAq})
and (\ref{2piDAg}) can be isospin decomposed and expressed in terms of
only two independent quark $2\pi$DA's corresponding to isospin $I=0$
and $I=1$ pair production respectively and one single gluon $2\pi$DA.
For $\pi^+\pi^-$ production we have
\be
\Phi^{\pi^+\pi^-}_{f'f}(z,\zeta,\mpp)&=&
\delta^{f'f} \Phi^{I=0}(z,\zeta,\mpp)
+\tau_3^{f'f} \Phi^{I=1}(z,\zeta,\mpp)\label{isodecomp1}\\
\Phi^{\pi^+\pi^-}_G(z,\zeta,\mpp)&=&\Phi^G(z,\zeta,\mpp)
\ee
and for $\pi^0\pi^0$ production
\be
\Phi^{\pi^0\pi^0}_{f'f}(z,\zeta,\mpp)&=&
\delta^{f'f} \Phi^{I=0}(z,\zeta,\mpp)\\
\Phi^{\pi^0\pi^0}_G(z,\zeta,\mpp)&=&\Phi^G(z,\zeta,\mpp)\,.
\label{isodecomp2}
\ee
Taking into account the $C$-invariance of the underlying theory one
can easily derive the following symmetry properties for the three
$2\pi$DA's:
\be
\Phi^{I=0}(z,\zeta,\mpp)&=&
-\Phi^{I=0}(1-z,\zeta,\mpp)
=\Phi^{I=0}(z,1-\zeta,\mpp)
\label{symmetry1}\\
\Phi^{I=1}(z,\zeta,\mpp)&=&\Phi^{I=1}(1-z,\zeta,\mpp)
=-\Phi^{I=1}(z,1-\zeta,\mpp)\label{symmetry2}
\\
\Phi^{G}(z,\zeta,\mpp)&=&
\Phi^{G}(1-z,\zeta,\mpp)
=\Phi^{G}(z,1-\zeta,\mpp)\label{symmetry3}
\ee
Using the decompositions (\ref{isodecomp1}) -- (\ref{isodecomp2}) and
the symmetry relations (\ref{symmetry1}), (\ref{symmetry2})
the transition amplitudes (\ref{Tmatrix}) for $\pi^+\pi^-$ and
$\pi^0\pi^0$ production and an elastically scattered proton target can
be written in the form
\be
T^{\pi^+\pi^-}&=&T^{I=0}+T^{I=1}\label{Tpm}\\
T^{\pi^0\pi^0}&=&T^{I=0}
\ee
with
\be
T^{I=0}&=&-(e4\pi\alphas)\frac{C_F}{N_c}
 \frac{1}{12Q}\,\tilde\Phi^{I=0}\label{TI0}
 \Bigl(2I_u^- -I_d^-\Bigr)
\\
T^{I=1}&=&-(e4\pi\alphas)\frac{C_F}{N_c}
 \frac{1}{12Q}\,\tilde\Phi^{I=1}
 \Bigl(2I_u^+ +I_d^+
+\frac{3}{C_F}I_G\Bigr)\,.\label{TI1}
\ee
Here $\tilde\Phi^{I=0}$ and $\tilde\Phi^{I=1}$ are functions of $\mpp$
and $\cos\theta$ defined as
\be
\tilde\Phi^{I=0}
&=& \int_0^1\d z
\frac{(1-2 z)\Phi^{I=0}(z,\zeta,\mpp)
-\frac{2}{C_F}\Phi^{G}(z,\zeta,\mpp)}{z(1-z)}
\label{Phi0int}\\
\tilde\Phi^{I=1}
&=& \int_0^1\d z
\frac{\Phi^{I=1}(z,\zeta,\mpp)}{z(1-z)}
\label{Phi1int}
\ee
and $I_f^\pm$, $I_G$ are the following integrals over SPD's depending
on $\xBj$ and $t$:
\be
I_f^\pm&=&\int_{-1}^1 \d\tau F_{ff}(\tau,\xi,t)\biggl[
\frac{1}{\tau+\xi-\i\epsilon} \pm \frac{1}{\tau-\xi+\i\epsilon}
\biggr]\label{Ipm}\\
I_G&=&\int_{-1}^1 \d\tau F_G(\tau,\xi,t) \biggl[
\frac{1}{\tau+\xi-\i\epsilon} + \frac{1}{\tau-\xi+\i\epsilon}
\biggr]
\ee

In the next section we will show that restricting the $z$-dependence
of the $2\pi$DA's to their asymptotic shapes but keeping the momentum
fraction $M_2^Q$ carried by quarks in the pion (gluons have the
momentum fraction $M_2^G=1-M_2^Q$) as a free parameter these functions
take the form
\be
\Phi^{I=0}(z,\zeta,\mpp)&=& -\frac{120 M_2^Q}{N_f}z(1-z)(2 z-1)
\biggl[ \frac{3C-\beta^2}{12}f_0(\mpp)P_0(\cos\theta)
-\frac{\beta^2}{6}f_2(\mpp)P_2(\cos\theta) \biggr]\\
\Phi^{I=1}(z,\zeta,\mpp)&=& 6 z(1-z) (2\zeta-1) F_\pi(\mpp)\\
\Phi^G(z,\zeta,\mpp)&=& - 60 M_2^G z^2(1-z)^2 \biggl[
\frac{3C-\beta^2}{12}f_0(\mpp)P_0(\cos\theta)
-\frac{\beta^2}{6}f_2(\mpp)P_2(\cos\theta) \biggr]\,, \ee where
$f_0$ and $f_2$ are Omn\`es functions, $F_\pi(\mpp)$ is the
electromagnetic form factor of the pion,
$N_f$ the number of light quark flavors, and
$C=1+bm_\pi^2$ with $b=-1.7\,{\mathrm GeV}^{-2}$ is an integration
constant estimated in the instanton model of the QCD vacuum \cite{MVP99}.
Using these expressions
the integrals (\ref{Phi0int}) and (\ref{Phi1int}) can be
evaluated:
\be
\tilde\Phi^{I=0} &=& \biggl(\frac{40M_2^Q}{N_f}+\frac{20M_2^G}{C_F}\biggr)
\biggl[\frac{3C-\beta^2}{12}f_0(\mpp)P_0(\cos\theta)
-\frac{\beta^2}{6}f_2(\mpp)P_2(\cos\theta)
\biggr]
\label{tildePhiI0}\\
\tilde\Phi^{I=1} &=& 6\beta F_\pi(\mpp)P_1(\cos\theta)\label{tildePhiI1}
\ee

The differential cross section of the electroproduction process is
given in terms of its $T$-matrix element $\tilde T^{\pi^a\pi^a}$ by
\be
\frac{\d\sigma^{\pi^a\pi^b}}{\d\xBj\d y\d t\,\d\mpp\,\d\cos\theta}
=
\frac{\mpp\beta}{8(4\pi)^5 p\cdot l}\sum_{S',s'}|\tilde T^{\pi^a\pi^b}|^2\,,
\label{crosssec}
\ee
where $l$ is the initial lepton momentum, $y$ is defined by
$y=p\cdot q/p\cdot l$ and corresponds to the energy loss of
the scattered lepton in the proton rest frame, and the sum is understood
to run over the spin polarizations of the scattered nucleon and
lepton. The $T$-matrix element of the electroproduction reaction
$e^-(l)+B_1(p)\to \pi\pi(q')+e^-(l-q)+B_2(p+\Delta)$
is related to the amplitude of the sub-process (\ref{subprocess})
calculated above by
\be
\tilde T^{\pi\pi}
=\bar u(l-q,s')\gamma_\mu u(l,s)\varepsilon_L^\mu\frac{-e}{Q^2}T^{\pi\pi}\,.
\ee
For the spin sum one gets
\be
\sum_{S's'}|\tilde T^{\pi\pi}|^2
=\frac{2e^2(1-y)}{p\cdot l\,\xBj y^3}\sum_{S'}|T^{\pi\pi}|^2\,.
\label{spinsum}
\ee

Using (\ref{crosssec}) and (\ref{spinsum}) the intensity densities defined in
(\ref{IDgeneral}) are obtained to be
\be
\langle P_l(\cos\theta)\rangle^{\pi\pi}=
\frac{\int\d\xBj\d y\d t\d\mpp\,\mpp\beta(1-y)y^{-3}\xBj^{-1}
\int_{-1}^1\d\cos\theta
P_l(\cos\theta)\sum_{S'}|T^{\pi\pi}|^2}
{\int\d\xBj\d y\d t\d\mpp\,\mpp\beta(1-y)y^{-3}\xBj^{-1}
\int_{-1}^1\d\cos\theta
\sum_{S'}|T^{\pi\pi}|^2}\,.\label{ID}
\ee
The $\cos\theta$-integrals can be solved analytically, they only
involve integrals over products of 3 Legendre polynomials as can be
seen comparing the previous equation with (\ref{Tpm}) -- (\ref{TI1}),
(\ref{tildePhiI0}), and (\ref{tildePhiI1}). The full expressions are rather
lengthy and are shown in the appendix. Non-vanishing results are obtained for
$\pi^+\pi^-$-production for $l=1,2,3,4$. Note that for $l=1$ and $l=3$
only the interference term $(T^{I=0})^*T^{I=1}+c.c.$ contributes to the
nominator of Eq.~(\ref{ID}). These two intensity densities are highly
sensitive on the production amplitude for isoscalar pion pairs, a
vanishing $I=0$ amplitude would imply that
$\langle P_1(\cos\theta)\rangle^{\pi^+\pi^-}$ and
$\langle P_3(\cos\theta)\rangle^{\pi^+\pi^-}$ are
zero. For $\pi^0\pi^0$-production we can predict that only the intensity
densities $\langle P_2(\cos\theta)\rangle^{\pi^0\pi^0}$ and
$\langle P_4(\cos\theta)\rangle^{\pi^0\pi^0}$ do not vanish.

We have mentioned above that our analysis is valid only up to $1/Q$ power
corrections, which are not necessarily negligible at realistic scales in
experiments like HERMES. Therefore it is desirable to define
quantities for which at least some power suppressed
contributions are absent. Indeed we can exclude the basically uncontrolled
contribution of transversally polarized photons taking an
appropriate linear combination of $\langle P_1(\cos\theta)\rangle^{\pi\pi}$
and $\langle P_3(\cos\theta)\rangle^{\pi\pi}$. The analysis of \cite{Sek}
shows that the combination
$\langle P_1(\cos\theta)\rangle^{\pi\pi}+
\sqrt{7/3}\langle P_3(\cos\theta)\rangle^{\pi\pi}$ projects out the
component of vanishing helicity of the final two pion state
(see also appendix B of \cite{MMS}). Assuming $s$-channel helicity
conservation we can conclude that only longitudinal
polarized photons contribute to this component.

\section{Two-pion distribution amplitudes}

The two-pion distribution amplitudes defined in (\ref{2piDAq}) and
(\ref{2piDAg}) describe the fragmentation of a pair of collinear
partons (quarks or gluons) into the final pion pair \cite{Ter}.
Some properties of the $2\pi$DA's were already presented in the
previous section (Eqs.~(\ref{isodecomp1}) -- (\ref{symmetry3})).
In this section we want to discuss further constraints for these
functions that allow us to construct realistic models for them.

Following \cite{MVP99,KMP99} we decompose both quark and gluon
$2\pi$DA's in conformal and partial waves (see
\cite{MVP99,KMP99}). For the quark $2\pi$DA's the decomposition reads
\be
\label{expansion}
\Phi^{I=0}(z,\zeta,\mpp)
&=&6z(1-z)
\sum_{\scriptstyle n=1 \atop \scriptstyle {\mathrm odd}}^\infty
\sum_{\scriptstyle l=0 \atop \scriptstyle {\mathrm even}}^{n+1}
B_{nl}^{I=0}(\mpp) C_n^{3/2}(2 z-1)
P_l(\beta^{-1}(2\zeta-1))\\
\label{expansion1}
\Phi^{I=1}(z,\zeta,\mpp)
&=&6z(1-z)
\sum_{\scriptstyle n=0 \atop \scriptstyle {\mathrm even}}^\infty
\sum_{\scriptstyle l=1 \atop \scriptstyle {\mathrm odd}}^{n+1}
B_{nl}^{I=1}(\mpp) C_n^{3/2}(2 z-1) P_l(\beta^{-1}(2\zeta-1))\,,
\ee
while for the gluon $2\pi$DA we have
\be
\label{expansion2}
\Phi^G(z,\zeta,\mpp)=
30z^2(1-z)^2
\sum_{\scriptstyle n=0 \atop \scriptstyle {\mathrm even}}^\infty
\sum_{\scriptstyle l=0 \atop \scriptstyle {\mathrm even}}^{n+2}
A^G_{nl}(\mpp) C^{5/2}_n(2 z-1) P_l(\beta^{-1}(2\zeta-1)),
\ee
with the Legendre polynomials $P_l(x)$ and the Gegenbauer polynomials
$C_n^\mu(x)$. The expansion of the $z$-dependence in Gegenbauer
polynomials is chosen such that under evolution of the $2\pi$DA's the
coefficients $B_{nl}^{I=0}$, $B_{nl}^{I=1}$, and $A^G_{nl}$ are
renormalized multiplicatively with mixing only between
$B_{nl}^{I=0}$ and $A^G_{n-1,l}$ \cite{ERBL,mix}.
With the choice of Legendre polynomials for the $\zeta$-expansion the
above decomposition is directly related to the partial wave expansion
of the resulting production amplitude for pion pairs.

The $2\pi$DA's are related to the distribution amplitudes of a single pion
$\varphi_\pi(z)$ by soft pion theorems of the form \cite{MVP99,KMP99}
\be
\Phi^{I=0}(z,\zeta=1,\mpp=0)&=&
\Phi^{I=0}(z,\zeta=0,\mpp=0)=0\label{spthI0}\\
\Phi^{I=1}(z,\zeta=1,\mpp=0)&=&
-\Phi^{I=1}(z,\zeta=0,\mpp=0)=\varphi(z)\\
\Phi^G (z,\zeta=1,\mpp=0)&=&
\Phi^G (z,\zeta=0,\mpp=0)=0\label{spthG}\,.
\ee
Additional constrains are provided by the
crossing relations between the quark and gluon $2\pi$DA's and the
corresponding (skewed and forward) parton distributions in the
pion. For the derivation of the crossing relations see \cite{MVP99}
(for quark $2\pi$DA's) and \cite{KMP99} (for the gluon
$2\pi$DA). Expressed in terms of the coefficients $B_{nl}^{I}$, and
$A^G_{nl}$ they take the form
\be
B_{n-1,n}^{I=0}(0)&=&
\frac{2}{3}\frac{2n+1}{n+1} \int_0^1 \d x\,x^{n-1}
\frac{1}{N_f}\sum_f(q^f_\pi(x)+q^{\bar f}_\pi(x))\\
B_{n-1,n}^{I=1}(0)&=&
\frac{2}{3}\frac{2n+1}{n+1} \int_0^1 \d x\,x^{n-1}
(q^u_{\pi^+}(x)-\bar q^{\bar u}_{\pi^+}(x))\\
A_{n-2,n}^G(0)&=&
\frac{4}{5}\frac{2n+1}{(n+1)(n+2)}
\int_0^1 \d x\,x^{n-1}
g_\pi(x)\,,
\ee
where $q^f_\pi(x)$, and $g_\pi(x)$ are the usual quark and gluon
distributions in the pion. Using these relations one
can easily derive the normalization of the $2\pi$DA's $\Phi^{I=0}$ and
$\Phi^G$ at $\mpp=0$:
\be
\int_0^1 \d z (2 z-1)\Phi^{I=0}(z,\zeta,\mpp=0)
&=&
-\frac{4}{N_f}\ M_2^Q\zeta(1-\zeta)\label{I0norm}\\
\int_0^1 \d z\Phi^G(z,\zeta,\mpp=0)
&=&
-2M_2^G\zeta(1-\zeta)\,.\label{Gnorm}
\ee
Here $M_2^Q$ and $M_2^G$ are the momentum fractions carried
by quarks or gluons respectively in the pion.
The first moment of $\Phi^{I=1}$ can be determined by crossing even
for arbitrary values of $\mpp$ and is expressed in terms of the
electromagnetic pion form factor  $F_\pi(\mpp)$ in the following way:
\be
\int_0^1 \d z \Phi^{I=1}(z,\zeta,\mpp)
&=&
(2\zeta-1) F_\pi(\mpp)\,.\label{I1norm}
\ee
This is due to the fact that crossing relates the moment on the left
hand side of Eq.~(\ref{I1norm}) to the first moment of the corresponding
pion SPD which is given by the pion form factor.

Before studying the $\mpp$-dependence of $\Phi^{I=0}$ and $\Phi^G$
we want to discuss the asymptotic shape of the $2\pi$DA's. In the
asymptotic limit only the coefficients
$B_{10}^{I=0}$, $B_{12}^{I=0}$, $B_{01}^{I=1}$, $A_{00}^{G}$, and
$A_{02}^{G}$ do not vanish. Therefore the asymptotic expression for
the isovector $2\pi$DA has the form \cite{PW98}
\be
\Phi^{I=1}(z,\zeta,\mpp)=
6z(1-z)(2\zeta-1)F_\pi(\mpp)\,,\label{PhiI1}
\ee
while taking into account the soft pion theorems (\ref{spthI0}),
(\ref{spthG}) and the normalization conditions (\ref{I0norm}),
(\ref{Gnorm}) we get for $\Phi^{I=0}$ and $\Phi^G$ at asymptotically
large $Q^2$ and vanishing mass $\mpp$
\be
\Phi^{I=0}(z,\zeta,\mpp =0)&=&
-\frac{120 M_2^Q}{N_f}z(1-z)(2z-1)\zeta(1-\zeta)\\
\Phi^G(z,\zeta,\mpp=0)&=&
-60 M_2^G z^2(1-z)^2\zeta(1-\zeta)\,.
\ee
The explicit form of the $Q^2$-dependence in LO implies that the
linear combinations
$A_{00}^{G}-C_F\frac{6}{5}B_{10}^{I=0}$ and
$A_{02}^{G}-C_F\frac{6}{5}B_{12}^{I=0}$ die out at large $Q^2$
due to the mixing between the gluon and the quark $2\pi$DA.
This can be used to fix the
asymptotic values for $M_2^Q$ and $M_2^G$ to
\be
M_2^{Q{\mathrm (asympt.)}}&=&\frac{N_f}{N_f+4C_F}\\
M_2^{G{\mathrm (asympt.)}}&=&\frac{4C_F}{N_f+4C_F}\,.
\ee
However, in order to be more general we do not restrict ourselves to
these values but keep $M_2^Q$ as a free parameter.

The $\mpp$-dependence of the $2\pi$DA's can be analyzed by means of
dispersion relations \cite{MVP99}. A similar analysis can be found
e.g.\ in Refs.~\cite{raby,gasser}, where the effect of final state
interactions on the decay of a light meson into two pions is
investigated. In the region $\mpp^2<16 m_\pi^2$ the imaginary part of
a $2\pi$DA is related to the pion--pion scattering amplitude due to
final state interactions by Watson's theorem \cite{watson}.
For $\Phi^{I=0}$ (In the analysis of $\Phi^G$ the line of argument is
exactly the same) this relation can be written in the form
\be
\mbox{Im}\{B_{nl}^{I=0}(\mpp)\}=
\sin(\delta_l^0(\mpp^2))
\e^{\i\delta_l^0(\mpp^2)}
B_{nl}^{I=0}(\mpp)^\ast=
\tan(\delta_l^0(\mpp^2))
\mbox{Re}\{B_{nl}^{I=0}(\mpp)\}\,,\label{imbnl}
\ee
where $\delta_l^I$ are the two pion phase shifts.
From Eq.~(\ref{imbnl}) the following $N$-fold subtracted dispersion
relation for the coefficients $B_{nl}^I(\mpp)$ can be derived:
\be
B_{nl}^{I=0}(\mpp)=\sum_{k=0}^{N-1}
\frac{\mpp^{2k}}{k!}\frac{\d^k}{\d\mpp^{2k}}
B_{nl}^{I=0}(0)+
\frac{\mpp^{2N}}{\pi}\int_{4m_\pi^2}^\infty
\d s \frac{\tan(\delta_l^0(s))\mbox{Re}\{B_{nl}^{I=0}(\sqrt{s})\}}
{s^N(s-\mpp^2-i\epsilon)}\,.
\ee
Solutions of such dispersion relations were found long ago
by  Omn\`es \cite{omnes} and can be written in the form\footnote{We
  give here the simplest form of the solution with only one
  subtraction and assuming that $B_{nl}^{I=0}(\mpp)$
  has no zeros in the relevant $\mpp$-interval. The
  presence of a zero in $B_{nl}^{I=0}(\mpp)$ at some value of $\mpp$
  can be interpreted as a signature for a glueball in the $\pi\pi$
  channel. This case shall be considered elsewhere.}
\be
\frac{B_{nl}^{I=0}(\mpp)}{B_{nl}^{I=0}(0)}=f_l(\mpp)=
\exp\Bigg[\i\delta_l^0(\mpp^2)+
\frac{\mpp^2}{\pi}
\mbox{Re}\bigg\{\int_{4m_\pi^2}^\infty
\d s \frac{\delta_l^0(s)}{s(s-\mpp^2-i\epsilon)}\bigg\}
\Bigg]\,,\label{omnessolution}
\ee
where $f_l(\mpp)$ are the so called Omn\`es functions.
By derivation, the solution (\ref{omnessolution}) is valid only for
$\mpp^2<16 m_\pi^2$, however, the deviations due to inelastic
pion--pion scattering are expected to be small up to rather high energies.
In terms of the Omn\`es functions the (quasi-) asymptotic
$2\pi$DA's $\Phi^{I=0}$ and $\Phi^G$ take the form
\be
\Phi^{I=0}(z,\zeta,\mpp)&=&
-\frac{120 M_2^Q}{N_f}z(1-z)(2 z-1)
\biggl[
\frac{3C-\beta^2}{12}f_0(\mpp)P_0(\cos\theta)
-\frac{\beta^2}{6}f_2(\mpp)P_2(\cos\theta)
\biggr]\label{PhiI0}\\
\Phi^G(z,\zeta,\mpp)&=&
- 60 M_2^G z^2(1-z)^2
\biggl[
\frac{3C-\beta^2}{12}f_0(\mpp)P_0(\cos\theta)
-\frac{\beta^2}{6}f_2(\mpp)P_2(\cos\theta)
\biggr]\,.\label{PhiG}
\ee
The constant $C$ in Eqs.~(\ref{PhiI0}) and (\ref{PhiG})
plays the role
of an integration constant in the Omn\`es solution of the
corresponding dispersion relation. From the soft pion theorem it
follows that $C=1+{\cal O}(m_\pi^2)$. Using the instanton model for
calculations of $B_{nl}(\mpp)$ at low energies
\cite{MVP99,KMP99} one finds the constant $C$ to be equal to:
\be
C=1+b m_\pi^2 + {\cal O}(m_\pi^4)\quad{\mathrm with}\quad b
\approx -1.7\,{\mathrm GeV}^{-2}\,.
\ee
In the section~\ref{echl} we shall discuss the relation of the
constant $C$ and the near threshold behavior of $f_0(\mpp)$ to the
effective chiral Lagrangian.
The expressions (\ref{PhiI1}), (\ref{PhiI0}), and (\ref{PhiG}) are
exactly those we have used for the $2\pi$DA's in the analysis of the
previous section.

\section{Modeling the skewed parton distributions}

The skewed parton distributions defined in Eqs.~(\ref{SPDq}) and
(\ref{SPDg}) can be decomposed into different Lorentz
structures. Adopting the notation of \cite{JiReview} we can write
\be
F_{ff}(\tau,\xi,t)&=&H_f(\tau,\xi,t)\,\bar u(p',S')\feynmandag n u(p,S)+
E_f(\tau,\xi,t)\,
\bar u(p',S')\frac{\i \sigma_{\mu\nu}n^\mu\Delta^\nu}{2m_N}u(p,S)\\
F_G(\tau,\xi,t)&=&H_G(\tau,\xi,t)\,\bar u(p',S')\feynmandag n u(p,S)+
E_G(\tau,\xi,t)\,
\bar u(p',S')\frac{\i \sigma_{\mu\nu}n^\mu\Delta^\nu}{2m_N}u(p,S)\,.
\ee
The SPD's $H_f(\tau,\xi,t)$ and $H_G(\tau,\xi,t)$ are highly
constrained by their forward limit whereas only little is known about
the functions $E_f(\tau,\xi,t)$ and $E_G(\tau,\xi,t)$ (however, it is
clear that their contribution vanishes when the skewedness $\xi$ goes
to zero).

For our analysis we model the functions $H_f(\tau,\xi,t)$ and
$H_G(\tau,\xi,t)$ using Radyushkin's double distributions \cite{Rad}.
The SPD's $H(\tau,\xi,t)$ are related to the so called
nonforward parton distributions ${\cal F}_{\zeta}(X,t)$ by:
\be
H_f(\tau,\xi,t)&=&\frac{1}{1+\xi}\biggl[\Theta(\tau+\xi)
{\cal F}^f_{\zeta}\Big(\frac{\tau+\xi}{1+\xi},t\Big)
-\Theta(\xi-\tau)
{\cal F}^{\bar f}_{\zeta}\Big(\frac{\xi-\tau}{1+\xi},t\Big)\biggr]\\
\tau H_G(\tau,\xi,t)&=&\frac{1}{2}\biggl[\Theta(\tau+\xi)
{\cal F}^G_{\zeta}\Big(\frac{\tau+\xi}{1+\xi},t\Big)
-\Theta(\xi-\tau)
{\cal F}^G_{\zeta}\Big(\frac{\xi-\tau}{1+\xi},t\Big)\biggr]
\ee
with $\zeta=\frac{2\xi}{1+\xi}$.
The skewed parton distributions are obtained from the double
distributions by the integral
\be
{\cal F}_{\zeta}(X,t)
=\int_0^{{\mathrm min}\{\frac{X}{\zeta},\frac{1-X}{1-\zeta}\}}
\d y F(X-\zeta y,y;t)\,.
\ee
For the double distributions we adopt now the ansatz suggested by
Radyushkin \cite{Rad99} $F(x,y;t=0)=q(x)\,\pi(x,y)$ (for quark
distributions) and $F(x,y;t=0)=x\,G(x)\,\pi(x,y)$ (for gluon
distributions), where $q(x)$ and $G(x)$ are the ordinary quark and
gluon parton distributions in the nucleon and $\pi(x,y)$ is a profile
function chosen to be $\pi(x,y)=6\,y(1-x-y)(1-x)^{-3}$ for quarks and
$\pi(x,y)=30\,y^2(1-x-y)^2(1-x)^{-5}$ for gluons. The (forward)
parton distributions are modeled using the MRS(A') parameterizations of
the reference \cite{MRS} at $Q^2=4\,{\mathrm GeV}^2$
(see footnote\,\footnote{We
  take the SPD's at this fixed scale. For evolution effects see
  e.g.\ \cite{evol}.}).
For the $t$-dependence of the SPD's we use a factorized ansatz
\cite{MPW,Vander} motivated by sum rules that relate their first
moments to form factors \cite{Ji}. The quark SPD's are constrained by
\be
\int_{-1}^1 \d\tau H_f(\tau,\xi,t) = F^{f/p}_1(t)\label{sumrule}
\ee
with the Pauli form factor $F_1$ of the proton with respect to the
flavor $f$. The corresponding ansatz for the symmetric part of
$H_f(\tau,\xi,t)$ is
\be
\frac{1}{2}\Bigl[H_f(\tau,\xi,t)+H_f(-\tau,\xi,t)\Bigr]
=\frac{1}{2}\Bigl[H_f(\tau,\xi,0)+H_f(-\tau,\xi,0)\Bigr]
\frac{F^{f/p}_1(t)}{F^{f/p}_1(0)}\,.
\ee
We adopt a parameterization for the nucleon form factors taken from the
appendix of \cite{Weise} and take into account the relations
\be
F_1^{u/p}=2 F_1^p+F_1^n\,,\quad
F_1^{d/p}=2 F_1^n+F_1^p\,,
\ee
which are valid if the contribution of strange (and heavier) quarks to the
nucleon form factor is neglected.
Taking the first moment of $\tau H_G(\tau,\xi,t)$ we can analogously
motivate the model
\be
H_G(\tau,\xi,t) = H_G(\tau,\xi,0)\frac{F_{\theta}(t)}{F_{\theta}(0)}
\ee
with the gluon form factor of the proton defined by
\be
\langle p'|G_{\mu\lambda}(0)G^\lambda_{\;\nu}(0)|p\rangle
=(2p_\mu p_\nu-\frac{1}{2}g_{\mu\nu}m_N^2)F_{\theta}(t)\,.
\ee
For the gluon form factor we can adopt the parameterization of
\cite{BGMS93}:
\be
F_{\theta}(t)=\frac{F_{\theta}(0)}{(1-t/(\alpha\mu^2))^\alpha}
\ee
with $\alpha=3$ and $\mu^2=2.6\,{\mathrm GeV}^2$.
The same $t$-dependence we assume also for the antisymmetric part of
$H_f(\tau,\xi,t)$ entering the integral $I_f^+$ (see
Eq.~(\ref{Ipm})). The combination $H_f(\tau,\xi,t)-H_f(-\tau,\xi,t)$
is not constrained by the sum rule (\ref{sumrule}), however, we expect
a behavior similar to the gluon SPD due to the fact that
$H_f(\tau,\xi,t)-H_f(-\tau,\xi,t)$ and $H_G(\tau,\xi,t)$ mix when they
are evolved.

In \cite{PW99} it was shown that the SPD's obtained by double
distributions are not complete and that an additional term has to be
added, the so called D-term. In our analysis we take into account this
result by adding to the Radyushkin model result for $H_f(\tau,\xi,t)$ a
term estimated from the chiral quark-soliton model.
We write
\be
H_f(\tau,\xi,0) = H_f^{{\mathrm (Radyushkin\ model)}}(\tau,\xi,0) +
\Theta(\xi-|\tau|)\frac{1}{N_f} D(\tau/\xi)
\ee
taking for the function $D(x)$ the following numerical estimate
\be
D(x)=-4\ (1-x^2)\Bigl[C_1^{3/2}(x) + 0.3\ C_3^{3/2}(x)
+0.1\ C_5^{3/2}(x)\Bigr]
\ee
extracted from a calculation of the singlet quark SPD
in the chiral quark-soliton model \cite{Petrov}.
The same term with the opposite sign and multiplied with the
corresponding form factor is taken as model for $E_f(\tau,\xi,t)$
in order to satisfy the sum rule \cite{JiReview}
\be
\int_{-1}^1 \d\tau\,\tau \sum_f
\Bigl[
H_f(\tau,\xi,t)+ E_f(\tau,\xi,t)\Bigr]=
{\mathrm independent\ of\ }\xi\,.
\label{Jisumrule}
\ee
(Note that the part of $H_f(\tau,\xi,t)$ resulting from a double
distribution automatically satisfies Eq.~(\ref{Jisumrule}).)

\section{Probing the effective chiral Lagrangian in hard exclusive
reactions}\label{echl}

In this section we consider the case when the produced pion pairs have
an invariant energy close to the threshold $\mpp=4m_\pi^2$.
In this case the dependence of the intensity densities on $\mpp$
is related to the effective chiral Lagrangian (EChL) describing
the interaction of soft pions with gravity. Therefore one can use data
on hard exclusive pion pair production to probe this yet unknown
part of the EChL.

Due to the QCD factorization theorem (\ref{factorization}) one has
a well defined separation of the short and large distance parts of
the interaction. Since the large distance parts of the process are
defined in terms of hadronic matrix elements of QCD quark and
gluon operators which are independent of the hard scale (up to a
logarithmic scale dependence which is controlled by well known
evolution equations) one can apply the methods of effective
chiral perturbation theory to describe properties of
these matrix elements in a model independent way.

In the analysis below we make the assumption that in hard
electroproduction of pion pairs the two pions are produced
dominantly by the operators of the lowest conformal
spin (index $n$ in Eqs.~(\ref{expansion}),
(\ref{expansion1}), and (\ref{expansion2})).
This assumption is justified for large $Q^2$ since the contribution of
operators with higher conformal spin logarithmically dies out with
increasing $Q^2$ due to their larger anomalous dimensions.
For the $I=1$ channel the lowest conformal spin corresponds to the vector
current operator and for $I=0$ to the energy momentum tensor.
Therefore using this assumption the constant $C$ in Eqs.~(\ref{PhiI0})
and (\ref{PhiG}) and the near threshold behavior of the functions
$F_\pi(\mpp)$, $f_0(\mpp)$, and $f_2(\mpp)$,
which enter the expressions for the various intensity densities
(see the appendix for a collection of formulae), are fixed by the EChL.

Let us write the relevant terms of
the fourth order EChL describing the interactions of soft pions with
electromagnetic fields and with gravity \cite{GL1,DL}:
\be
\label{kirlag}
{\cal L}^{(4)}=-\i L_9{\mathrm Tr}\Big[
F^{\mu\nu} \partial_\mu U\partial_\nu U^\dagger \Big]+
L_{11} R\,{\mathrm Tr} \Big[ \partial_\mu U \partial^\mu U^\dagger
\Big]+
L_{12} R_{\mu\nu}{\mathrm Tr} \Big[ \partial^\mu U \partial^\nu U^\dagger
\Big]+
L_{13} R\ {\mathrm Tr} \Big[ m U + U^\dagger m
\Big]+ \ldots
\ee
Here $F^{\mu\nu}$ is the field strength of the photon field, $R_{\mu\nu}$
and $R$ are the Ricci tensor and the curvature scalar of an external
gravitational field, and $U=exp(i \pi^a \lambda^a/f_\pi)$ is the
non-linear pseudo-Goldstone field. The ellipsis stands for the
terms of the EChL that are not relevant for us here.
Now we can use the results of the calculations in Refs.~\cite{GL1,DL} to
express the constant $C$ and the near threshold behavior of the functions
$F_\pi$, $f_0$, and $f_2$ in terms of the constants $L_i$ in the EChL
(\ref{kirlag}). Let us first introduce the scale dependent
constants $L_i^r(\mu)$ which appear after proper renormalization
of ultra-violet divergences we have
\be
&&L^r_i(\mu)=L_i - \Gamma_i\;\lambda\,,\\
&&\Gamma_9=\frac{1}{4},\ \Gamma_{11}=\frac{1}{4}, \ \Gamma_{12}=0,\
\Gamma_{13}=\frac{2}{9}\,,\\
&&\lambda=\frac{\mu^{d-4}}{(4\pi)^2}\biggl[
\frac{1}{d-4} -\frac 12 \Bigl(\ln 4\pi +\Gamma'(1)+1\Bigr) \biggr]\,,
\ee
for further details see Refs.~\cite{GL1,DL}.
Using these definitions and the results of chiral perturbation theory
\cite{GL1,GL2} we obtain for $F_\pi(\mpp)$
\be
F_\pi(\mpp)
=1+\frac{2\mpp^2}{f_\pi^2} L_9^r(\mu) +
\frac{4 m_\pi^2\!-\mpp^2}{96\pi^2} K_\pi(\mpp)+\frac{4 m_K^2\!-\mpp^2}
{192\pi^2}
K_K(\mpp)
-\frac{\mpp^2}{192 \pi^2}\biggl(
1\!+\!2 \ln\frac{m_\pi^2}{\mu^2}\!+\!\ln\frac{m_K^2}{\mu^2}
\biggr)\,,
\ee
where the standard loop integral is defined by
\be
K_P(\mpp)=\int_0^1 \d x\;\ln\bigg(
1-x(1-x)\frac{\mpp^2}{m_P^2}\bigg)\,.
\ee
The result for the constant $C$ and the near threshold behavior
of $f_0(\mpp)$ and $f_2(\mpp)$
can be easily extracted from the results of Ref.~\cite{DL}:
\be
C&=&1- \frac{ m_\pi^2}{f_\pi^2} \biggl[
8\Bigl(4 L_{11}^r(\mu)+L_{12}^r(\mu)
-2L_{13}^r(\mu)\Bigr)+ \frac{1}{48\pi^2}
\Bigl(5\ln\frac{\mu^2}{m_\pi^2}-
\frac{1}{3}\ln\frac{\mu^2}{m_\eta^2}
+2\ln\frac{\mu^2}{m_K^2}
-\frac {106}{15} \Bigr) \biggr]\\
C\,f_0(\mpp)&+&\frac{\beta^2}{3}\Bigl(f_2(\mpp)-f_0(\mpp)\Bigr)
=1+2\frac{\mpp^2}{f_\pi^2} \Bigl(4
L_{11}^r(\mu)+L_{12}^r(\mu)\Bigr)-16
\frac{m_\pi^2}{f_\pi^2}\Bigl(L_{11}^r(\mu)-L_{13}^r(\mu)\Bigr)\nn\\
&&-\frac{2\mpp^2-m_\pi^2}{16\pi^2 f_\pi^2} \bar K_\pi(\mpp)-
\frac{\mpp^2}{16\pi^2 f_\pi^2} \bar
K_K(\mpp)-\frac{m_\pi^2}{48\pi^2 f_\pi^2}\bar K_\eta(\mpp)\\
f_2(\mpp)&=&1-2 L_{12}^r(\mu) \frac{\mpp^2}{f_\pi^2}
\ee
Here the function $\bar K_P(\mpp)$ is defined by
\be
\bar K_P(\mpp) =\frac{2 m_P^2 +\mpp^2}{3 \mpp^2}
K_P(\mpp)+ \frac{1}{3} \Bigl(\ln\frac{m_P^2}{\mu^2}+\frac{4}{3}\Bigr)\,.
\ee

One sees that hard exclusive production of $\pi\pi$ pairs
at invariant masses close to the threshold $\mpp=2m_\pi$
opens a possibility to measure the coefficients $L_{11-13}$ of
the EChL. These coefficients
cannot be accessed experimentally in low energy processes as for
such experiments one would need a strong source of gravity.
Let us note that the couplings in the EChL involving gravity are also of
relevance for the description of a meson gas out of thermal
equilibrium \cite{nicola} (as e.g.\ in heavy ion collisions) as well as for
the decay of a hypothetical light Higgs meson \cite{meissner}. This simple
example shows that hard exclusive processes have a big potential
for studies of new properties of the EChL.
In general, this kind of hard reactions allows to create low
energy probes with quantum numbers that are not provided by
nature or are hardly accessible.

\section{Results and Discussion}

In section \ref{section2} we have deduced analytical expressions for
the cross sections and intensity densities in terms of skewed parton
distributions of the nucleon, the electromagnetic pion form factor,
the so called Omn\`es functions $f_0(\mpp)$ and $f_2(\mpp)$, and the
momentum fraction $M_2^Q$ carried by the quark in the pion.

The results of section \ref{section2} show that $M_2^Q$ enters the
expressions for cross sections and intensity densities only in the
combination $40M_2^Q/N_f+20M_2^G/C_F=20/C_F+(40/N_f-20/C_F)M_2^Q$.
For $N_f=3$ this combination is only weakly dependent on the value of
$M_2^Q$. Therefore we choose for the following analysis the asymptotic
expression $M_2^Q=N_f/(N_f+4C_F)$ and state that our predictions are
insensitive to the exact value of this parameter reducing in this way
the number of parameters of our result\footnote{Note that the process
$\gamma^*\gamma\to\pi\pi$, which is also sensitive to
gluon $2\pi$DA's (see Ref.~\cite{KMP99}), depends strongly on the ratio
$M_2^Q/M_2^G$ as noted recently in Ref.~\cite{KM00}}.
Let us note that for this
choice the contribution of the gluon $2\pi$DA to the isoscalar
amplitude is exactly twice as large as the contribution from the quark
$2\pi$DA.

\begin{figure}
\centering
\epsfig{file=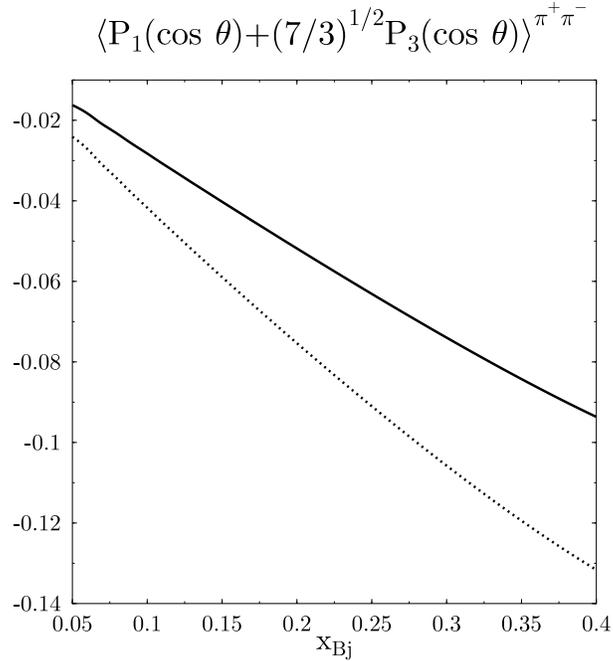, width=10cm}
\caption{$\langle P_1(\cos\theta)\rangle^{\pi^+\pi^-}+\sqrt{\frac{7}{3}}
  \langle P_3(\cos\theta)\rangle^{\pi^+\pi^-}$ as a function of
  $\xBj$ with cross sections integrated over $\mpp$ from the threshold
  to $0.6\,{\mathrm GeV}$. The dotted line shows the corresponding
  result obtained using the fit of [\ref{gasser}] instead of the
  Pad\'e approximation for the Omn\`es function $f_0(\mpp)$.}
\label{xplot}
\end{figure}

For the numerical results shown in this section we use the models for
the SPD's of the previous section. The pion form factor $F_\pi(\mpp)$
entering the isovector $2\pi$DA is modeled using its parameterization
given in \cite{guerro}, which is in very good agreement with
experimental data \cite{barkov}.
The Omn\`es functions contain information about the $\pi\pi$
resonances as well as about the non-resonant background.
The function $f_2(\mpp)$ is dominated by the $f_2(1270)$ resonance
resulting in a peak at $\mpp=1.275\,{\mathrm GeV}$.
In our analysis we use the expression (\ref{omnessolution}) for $f_2(\mpp)$
with the D-wave pion scattering phase shift $\delta_2^0$ given by
the Breit--Wigner contribution of the $f_2(1270)$ resonance.
For the the Omn\`es function $f_0(\mpp)$ we use (except in
Fig.~\ref{xplot}) the Pad\'e approximation suggested in \cite{f0}.

\begin{figure}
\centering
\epsfig{file=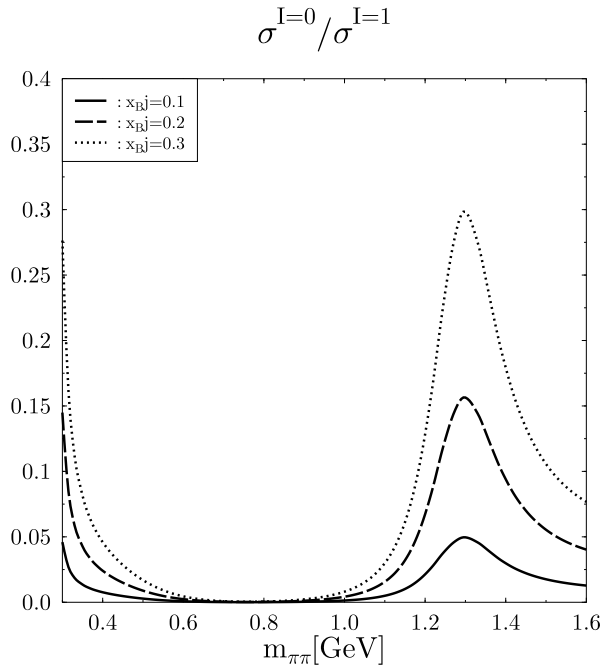, width=10cm}
\caption{The ratio of the differential cross sections for isoscalar
  and isovector pion pair production at three different values for $\xBj$
  as a function of $\mpp$.}
\label{ratio1}
\end{figure}

As a first result we show in Fig.~\ref{ratio1} the ratio of the
differential cross sections for the production of pion pairs with isospin
$I=0$ and $I=1$ as a function of the invariant mass of the two pions
$\mpp$ at the three different $\xBj$-values $\xBj=0.1$, $\xBj=0.2$,
and $\xBj=0.3$. The squared momentum transfer $t$ has been integrated
from $-0.6\,{\mathrm GeV}^2$ to $-t_{{\mathrm min}}=-m_N^2\xBj^2/(1-\xBj)$.
The plot shows that relatively large $I=0$ cross sections are to be
expected at small $\mpp$ close to the threshold due to the S-wave
contribution and around $1.3\,{\mathrm GeV}$ due to D-waves, which are
dominated by the $f_2(1270)$ resonance. Near the $\rho$
resonance around $0.77\,{\mathrm GeV}$, however, the isoscalar
contribution is negligible compared to the production of isovector
pion pairs.
The slight difference to our result in \cite{maxben} is due to the
inclusion of the gluon SPD, which contributes only to isovector pion
pair production, as well as to the fact that we have integrated over a
finite $t$-interval.

Figure~\ref{ratio1} shows that the most favorable kinematic
region to observe the isoscalar channel by the measurement of
intensity densities in $\pi^+\pi^-$ production is in the two $\mpp$
regions above or below the $\rho$ resonance and at relatively large
values of the Bjorken variable $\xBj$.

\begin{figure}
\centering
\epsfig{file=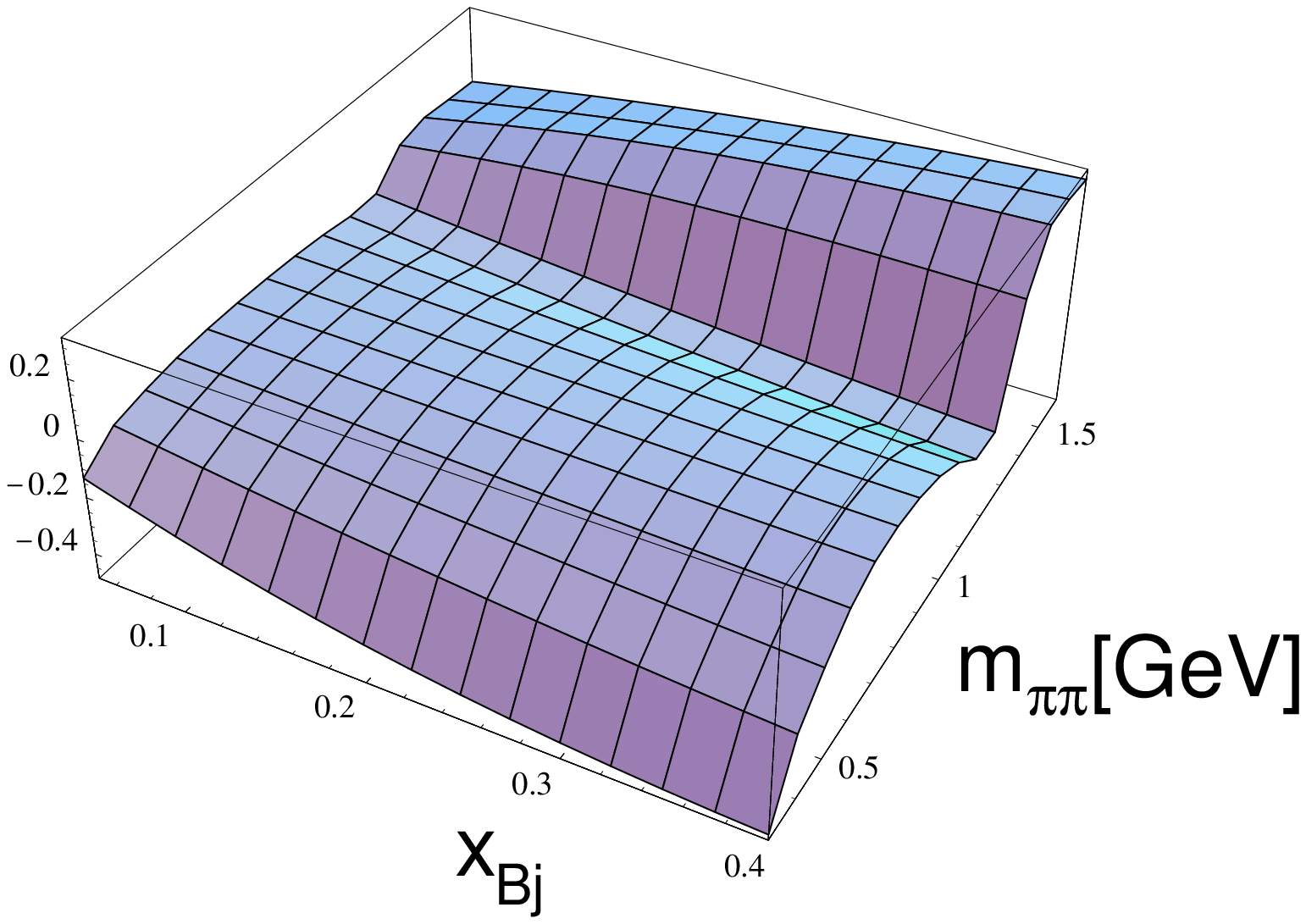, width=10cm}
\caption{$\langle P_1(\cos\theta)\rangle^{\pi^+\pi^-}$
  as a function of $\xBj$ and $\mpp$.}
\label{ID1Plot3D}
\end{figure}

\begin{figure}
\centering
\epsfig{file=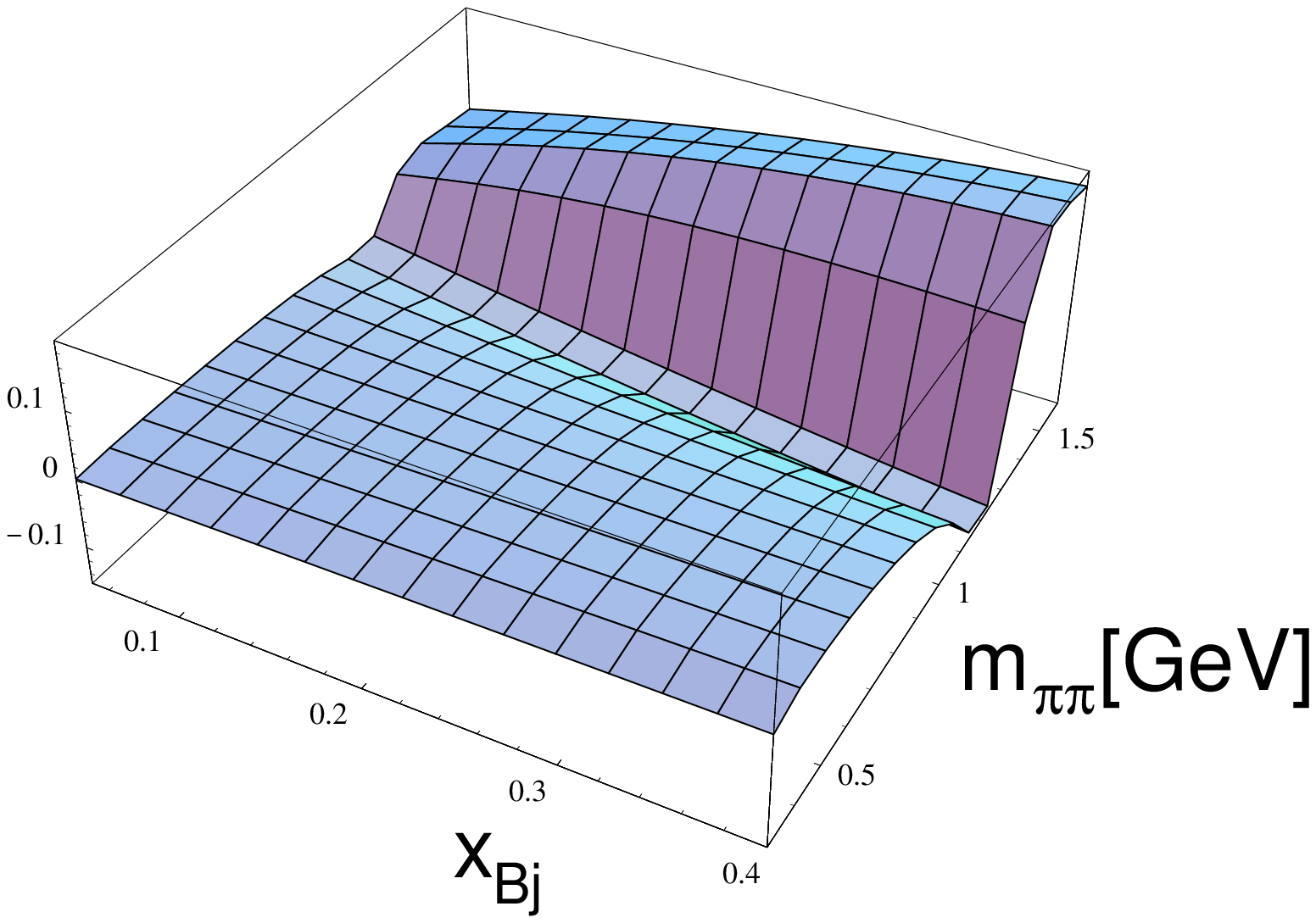, width=10cm}
\caption{$\langle P_3(\cos\theta)\rangle^{\pi^+\pi^-}$
  as a function of $\xBj$ and $\mpp$.}
\label{ID3Plot3D}
\end{figure}

This can also be seen from Figs.~\ref{ID1Plot3D} and \ref{ID3Plot3D},
where we show our results for the intensity densities
$\langle P_1(\cos\theta)\rangle^{\pi^+\pi^-}$ and
$\langle P_3(\cos\theta)\rangle^{\pi^+\pi^-}$ on the basis of
$t$-integrated cross sections as functions of $\xBj$ and
$\mpp$. The integration interval for the variable $t$ is again as in
the following figures $-0.6\,{\mathrm GeV}^2<t<-t_{{\mathrm min}}$.
Figure~\ref{ID3Plot3D} shows that
$\langle P_3(\cos\theta)\rangle^{\pi^+\pi^-}$ is sizable only in
the $f_2(1270)$ resonance region. This results from the fact that this
intensity density is proportional to the Omn\`es function $f_2(\mpp)$
as can be seen in Eq.~(\ref{A8}) in the appendix.

In Figs.~\ref{mplot} and \ref{xplot} we show our results for the
combination of intensity densities
$\langle P_1(\cos\theta)\rangle^{\pi^+\pi^-}+\sqrt{7/3}
\langle P_3(\cos\theta)\rangle^{\pi^+\pi^-}$ for $t$- and $\xBj$- or
$t$- and $\mpp$-integrated cross sections respectively as functions of
$\mpp$ or $\xBj$. The contribution of
$\langle P_3(\cos\theta)\rangle^{\pi^+\pi^-}$ in this combination is
relatively small, especially in the region of small $\mpp$-values near
the threshold. The choice of the kinematic region is motivated by
the above results and by the kinematic range of the HERMES experiment,
where corresponding measurements can be done \cite{HERMES}.

Let us mention that the expressions for the cross sections entering
our results for the intensity densities depend on the variable
$y=p\cdot q/p\cdot l$ only through the common factor
$(1-y)/y^3$ in Eq.~(\ref{spinsum}) as long as we neglect the scale
dependence of the SPD's. Therefore for comparison with experiments the
cross sections can be taken as integrated over an arbitrary
$y$-interval, a realistic choice could be $0.35<y<0.6$.

Especially in the region near the threshold $\mpp=0.28\,{\mathrm GeV}$
the predictions are very sensitive to the Omn\`es function
$f_0(\mpp)$. To illustrate this dependence we also show in
Fig.~\ref{xplot} the result for the alternative parameterization
of \cite{gasser} for $f_0(\mpp)$ (dotted line). We see that the
process of {\it hard} exclusive pion pair production is sensitive to the
mechanisms of the {\it low-energy} scattering of pions in the isoscalar
channel and therefore can be used to obtain new information
on the chiral dynamics in this channel.

We have checked that ignoring the D-term contribution to the
quark SPD changes the results by about 10\%. This may be taken as an
estimate for the uncertainty of our predictions due to a lack of
knowledge about the SPD's, especially about the functions $E(\tau,\xi,t)$.
Let us also note that NLO corrections in hard exclusive reactions,
which are not included in the analysis of the present article, can be
noticeable \cite{KMP99,NLO}.

So far we have restricted ourselves to electron--proton
scattering. The corresponding results for a neutron target can easily be
obtained by the interchange $I_u^-\leftrightarrow I_d^-$ and
$I_u^+\leftrightarrow I_d^+$ in Eqs.~(\ref{TI0}) and (\ref{TI1}) due
to isospin symmetry. A detailed analysis shows that in this case the
ratio of the amplitudes for isoscalar and isovector pion pair production is
approximately one order of magnitude smaller than for a proton target. This
is due to a cancelation of the contributions of u- and d-quark

SPD's in Eq.~(\ref{TI0}) because the u-quark density in the
neutron (corresponding to the d-quark density in the proton) is
roughly half as large as the d-quark density. Consequently we can
predict that also the intensity densities
$\langle P_1(\cos\theta)\rangle^{\pi^+\pi^-}$ and
$\langle P_3(\cos\theta)\rangle^{\pi^+\pi^-}$ are about one order
of magnitude smaller for neutron targets. This observation might help
to check the whole picture in experiments with deuteron targets. Such
experiments could also help to disentangle the contributions of the
different quark and gluon SPD's.

\begin{figure}
\centering
\epsfig{file=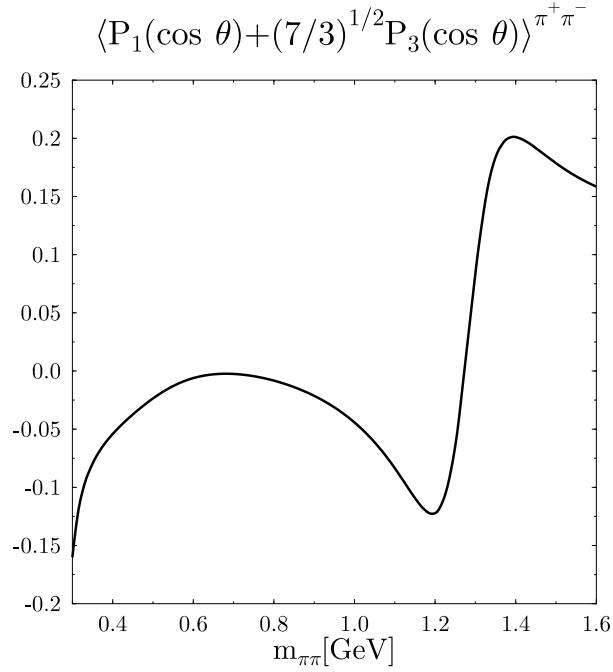, width=10cm}
\caption{$\langle P_1(\cos\theta)\rangle^{\pi^+\pi^-}+\sqrt{\frac{7}{3}}
    \langle P_3(\cos\theta)\rangle^{\pi^+\pi^-}$ as a
    function of $\mpp$ with cross sections integrated over $\xBj$
    from $0.05$ to $0.4$.}
\label{mplot}
\end{figure}

\section{Summary}

We have analyzed the angular distributions for exclusive electroproduction
of pion pairs. Our analysis was based on the amplitudes at leading
order in $1/Q^2$ and $\alphas$. These amplitudes
were expressed in terms of skewed parton distributions and two-pion
distribution amplitudes, for which we have used realistic models.
We have introduced intensity densities as Legendre moments of the
two-pion angular distributions and we have shown that they are a good probe
for the contribution of isoscalar pion pairs to the considered
process. Our analytic results show that isoscalar pion pairs are
mostly produced by two collinear gluons, which, in principle, opens a
possibility to study the gluon content of the isoscalar $\pi\pi$ states.
We predict sizable effects that could be measured at
experiments like HERMES. As a promising kinematic range for
a corresponding experiments we identified the $\mpp$-regions near the
threshold and around the $f_2(1270)$ resonance and relatively
large values of the Bjorken variable $\xBj>0.2$. The $\mpp$-dependence
should show a highly characteristic pattern. Our results are nearly
parameter free such that experimental data would provide an excellent
check for the whole formalism describing exclusive electroproduction
processes with skewed parton distributions. In addition we have shown that
studies of hard exclusive pion pair production near to the threshold
open a new way to probe the effective chiral Lagrangian.

\section*{Acknowledgments}

We are grateful to M.~Amarian,
H.~Avakian, B.~Clerbaux, L.~Frankfurt, A.~Kirchner,
N.~Kivel, L.~Mankiewicz, U.G.~Mei{\ss}ner, D.~M\"uller, P.V.~Pobylitsa,
S.~Schaefer, E.~Stein, M.~Strikman, and O.~Teryaev for
useful discussions. M.V.P. thanks J.~Gasser for providing the details of
Ref.~\cite{gasser}.  The work has been supported by the Russian-German
exchange program, BMBF, DFG, COSY-J\"ulich, and Studienstiftung des
deutschen Volkes.

\appendix
\section*{}

In this appendix are shown the full expressions for the
$\cos\theta$-integrated absolute squares of the $T$-matrix elements,
weighted with Legendre polynomials $P_l(\cos\theta)$, as they enter
the intensity densities.

Using Eqs.~(\ref{Tpm}) -- (\ref{TI1}), (\ref{tildePhiI0}), and
(\ref{tildePhiI1}) we get
\be
\int_{-1}^1\d\cos\theta\sum_{S'}|T^{\pi^+\pi^-}|^2
&=&\int_{-1}^1\d\cos\theta\sum_{S'}\Bigl(|T^{I=0}|^2+|T^{I=1}|^2\Bigr)\\
\int_{-1}^1\d\cos\theta\sum_{S'}|T^{\pi^0\pi^0}|^2
&=&\int_{-1}^1\d\cos\theta\sum_{S'}|T^{I=0}|^2
\ee
with
\be
\int_{-1}^1\d\cos\theta\sum_{S'}|T^{I=0}|^2
&=&
\frac{(e\pi\alphas)^2C_F^2}{18N_c^2}\frac{1}{p\cdot l\,\xBj y}
\sum_{S'}\Bigl|2I_u^--I_d^-\Bigr|^2
\biggl(\frac{40M_2^Q}{N_f}+\frac{20M_2^G}{C_F}\biggr)^2\nn\\
&&\times\Biggl[\frac{(3C-\beta^2)^2}{72}|f_0(\mpp)|^2
+\frac{\beta^4}{90}|f_2(\mpp)|^2\Biggr]
\ee
and
\be
\int_{-1}^1\d\cos\theta\sum_{S'}|T^{I=1}|^2
&=&
\frac{(e\pi\alphas)^2C_F^2}{18N_c^2}\frac{1}{p\cdot l\,\xBj y}
\sum_{S'}\Bigl|2I_u^++I_d^++\frac{3}{C_F}I_G\Bigr|^2
24\beta^2|F_\pi(\mpp)|^2\,.
\ee
The other non-vanishing integrals are:
\be
\int_{-1}^1\d\cos\theta\sum_{S'}|T^{\pi^+\pi^-}|^2P_1(\cos\theta)
&=&
\frac{(e\pi\alphas)^2C_F^2}{18N_c^2}\frac{1}{p\cdot l\,\xBj y}
\biggl(\frac{40M_2^Q}{N_f}+\frac{20M_2^G}{C_F}\biggr)\nn\\
&&\times
2{\mathrm Re}\Bigg\{\sum_{S'}
\Bigl(2I_u^--I_d^-\Bigr)^*
\Bigl(2I_u^++I_d^++\frac{3}{C_F}I_G\Bigr)\nn\\
&&\qquad\times
\Biggl[\frac{3C-\beta^2}{3}f_0(\mpp)^*
-\frac{4\beta^2}{15}f_2(\mpp)^*\Biggr]\beta F_\pi(\mpp)\Bigg\}\\
\int_{-1}^1\d\cos\theta\sum_{S'}|T^{\pi^+\pi^-}|^2P_2(\cos\theta)
&=&
\frac{(e\pi\alphas)^2C_F^2}{18N_c^2}\frac{1}{p\cdot l\,\xBj y}\nn\\
&&\times
\Biggl\{\sum_{S'}\Bigl|2I_u^--I_d^-\Bigr|^2
\biggl(\frac{40M_2^Q}{N_f}+\frac{20M_2^G}{C_F}\biggr)^2\nn\\
&&\qquad\times
\Biggl[\frac{3C-\beta^2}{90}\beta^2
{\mathrm Re}\Big\{f_0(\mpp)^*f_2(\mpp)\Big\}
+\frac{\beta^4}{315}|f_2(\mpp)|^2\Biggr]\nn\\
&&\quad+
\sum_{S'}\Bigl|2I_u^++I_d^++\frac{3}{C_F}I_G\Bigr|^2
\frac{144\beta^2}{35}|F_\pi(\mpp)|^2\Biggr\}\\
\int_{-1}^1\d\cos\theta\sum_{S'}|T^{\pi^0\pi^0}|^2P_2(\cos\theta)
&=&
\frac{(e\pi\alphas)^2C_F^2}{18N_c^2}\frac{1}{p\cdot l\,\xBj y}
\sum_{S'}\Bigl|2I_u^--I_d^-\Bigr|^2
\biggl(\frac{40M_2^Q}{N_f}+\frac{20M_2^G}{C_F}\biggr)^2\nn\\
&&\times
\Biggl[\frac{3C-\beta^2}{90}\beta^2
{\mathrm Re}\Big\{f_0(\mpp)^*f_2(\mpp)\Big\}
+\frac{\beta^4}{315}|f_2(\mpp)|^2\Biggr]\\
\int_{-1}^1\d\cos\theta\sum_{S'}|T^{\pi^+\pi^-}|^2P_3(\cos\theta)
&=&
-\frac{(e\pi\alphas)^2C_F^2}{18N_c^2}\frac{1}{p\cdot l\,\xBj y}
\biggl(\frac{40M_2^Q}{N_f}+\frac{20M_2^G}{C_F}\biggr)\nn\\
&&\times
2{\mathrm Re}\Bigg\{\sum_{S'}
\Bigl(2I_u^--I_d^-\Bigr)^*
\Bigl(2I_u^++I_d^++\frac{3}{C_F}I_G\Bigr)
\frac{6\beta^3}{35}f_2(\mpp)^*F_\pi(\mpp)\Bigg\}
\label{A8}\\
\int_{-1}^1\d\cos\theta\sum_{S'}|T^{\pi^+\pi^-}|^2P_4(\cos\theta)
&=&
\int_{-1}^1\d\cos\theta\sum_{S'}|T^{\pi^0\pi^0}|^2P_4(\cos\theta)\nn\\
&=&
\frac{(e\pi\alphas)^2C_F^2}{18N_c^2}\frac{1}{p\cdot l\,\xBj y}
\sum_{S'}\Bigl|2I_u^--I_d^-\Bigr|^2
\biggl(\frac{40M_2^Q}{N_f}+\frac{20M_2^G}{C_F}\biggr)^2
\frac{\beta^4}{315}|f_2(\mpp)|^2\,.
\ee

\end{document}